\documentclass[AMA,Times1COL]{WileyNJDv5} %STIX1COL,STIX2COL,STIXSMALL

\pdfoutput=1

\usepackage{subfigure}
\usepackage{color}
\newcommand{\infoset}{\mathcal{I}}
\newcommand{\cond}{\,\vert\,}

\usepackage{footmisc}

\articletype{Article Type}%

\received{Date Month Year}
\revised{Date Month Year}
\accepted{Date Month Year}
\journal{Journal}
\volume{00}
\copyyear{2023}
\startpage{1}

\begin{document}

\title{Combining Counterfactual Regret Minimization with Information Gain to Solve Extensive Games with Unknown Environments}

\author[1]{Chen Qiu}

\author[1,2]{Xuan Wang}

\author[1]{Tianzi Ma}

\author[1]{Yaojun Wen}

\author[1,2]{Jiajia Zhang}

\authormark{Chen Qiu \textsc{et al.}}
\titlemark{Combining Counterfactual Regret Minimization with Information Gain to Solve Extensive Games with Unknown Environments}

% \address[1]{Anonymous}
\address[1]{\orgdiv{School of Computer Science and Technology}, \orgname{Harbin Institute of Technology Shenzhen}, \orgaddress{\state{Shenzhen}, \country{China}}}

\address[2]{\orgdiv{Guangdong Provincial Key Laboratory of Novel Security Intelligence Technologies}, \orgaddress{\state{Shenzhen}, \country{China}}}

%%%%%%%%%%%%%%%%%%%%%%%%%%%%%%%%%%%%%%%%%%%
% \address[3]{\orgdiv{Department Name}, \orgname{Institution Name}, \orgaddress{\state{State Name}, \country{Country Name}}}
%%%%%%%%%%%%%%%%%%%%%%%%%%%%%%%%%%%%%%%%%%%

\corres{Xuan Wang, School of Computer Science and Technology, Harbin Institute of Technology Shenzhen, Shenzhen 518055, China, \email{wangxuan@cs.hitsz.edu.cn}}

% \presentaddress{This is sample for present address text this is sample for present address text.}

%\fundingInfo{Text}
%\JELinfo{ejlje}

\abstract[Abstract]{Counterfactual regret minimization (CFR) is an effective algorithm for solving extensive games with imperfect information (IIEGs). However, CFR is only allowed to be applied in known environments, where the transition function of the chance player and the reward function of the terminal node in IIEGs are known. In uncertain situations, such as reinforcement learning (RL) problems, CFR is not applicable. Thus, applying CFR in unknown environments is a significant challenge that can also address some difficulties in the real world. Currently, advanced solutions require more interactions with the environment and are limited by large single-sampling variances to narrow the gap with the real environment. In this paper, we propose a method that combines CFR with information gain to compute the Nash equilibrium (NE) of IIEGs with unknown environments. We use a curiosity-driven approach to explore unknown environments and minimize the discrepancy between uncertain and real environments. Additionally, by incorporating information into the reward, the average strategy calculated by CFR can be directly implemented as the interaction policy with the environment, thereby improving the exploration efficiency of our method in uncertain environments. Through experiments on standard testbeds such as Kuhn poker and Leduc poker, our method significantly reduces the number of interactions with the environment compared to the different baselines and computes a more accurate approximate NE within the same number of interaction rounds.}


\maketitle

% \renewcommand\thefootnote{}
% \footnotetext{\textbf{Abbreviations:} NE, Nash equilibrium; CFR, counterfactual regret minimization; IIEGs, extensive-form games with imperfect information.}

\footnotetext{A preprint has previously been published \cite{qiu2021combining}.}

\section{Introduction}\label{sec1}

Nash equilibrium (NE) \cite{nash1950equilibrium,DBLP:journals/cacm/Roughgarden10} is a significant concept for solving two-player extensive-form games with imperfect information (IIEGs) \cite{osborne1994course}. \emph{Counterfactual regret minimization} (CFR) \cite{zinkevich2007regret} and its variant algorithms \cite{gibson2012generalized,brown2018superhuman} are widely used for computing approximate NE. These algorithms require sufficient environmental information to be effectively applied. In contrast, reinforcement learning (RL) \cite{sutton2018reinforcement} algorithms provide strong assistance for agents to make decisions in uncertain environments by allowing them to interact with and acquire information from their surroundings, which in turn encourages agents to improve their strategies. RL methods enable agents to enhance their performance by setting suitable reward functions, although they may sometimes get trapped in local optimal states. Therefore, it is worth studying how to use CFR-like algorithms to find NE and design effective environment exploration strategies in IIEGs with unknown environments.

In this paper, we focus on 2-player zero-sum extensive-form games with imperfect information in unknown environments. To address this problem, Zhou et al. \cite{DBLP:conf/iclr/ZhouLZ20} proposed a state-of-the-art method called PSRLCFR. PSRLCFR employs the concept of \emph{thompson sampling} to model the environment, with the objective of narrowing the gap between the sampled and actual environments through exploration, primarily directed towards the most significant disparities between these two environments. Nonetheless, a notable limitation of PSRLCFR's exploration strategy arises from the substantial variance associated with single sampling, preventing accurate representation of the real environmental distribution. Furthermore, the exploration approach in PSRLCFR is influenced by the interaction strategy, necessitating a higher number of interactions between the agent and the environment for convergence to the actual environment. As a consequence, this method demands more iterative rounds throughout the entirety of the solution process to meet the requirements for finding a NE.

Effective exploration methods and interaction strategies play pivotal roles in addressing challenges of this nature. In this paper, we adopt \emph{variational information maximizing exploration} (VIME) \cite{houthooft2016vime} to improve the exploration efficiency of the environment and accelerate the speed of equilibrium convergence. VIME is a curiosity-driven exploration method that provides a novel approach to exploring unknown RL environments. Its dynamic model harnesses the information gain from the agent's internal belief as the propulsive force for exploration. This algorithm uniquely treats the information possessed by the agent as an integral part of the state, and its primary objective is to acquire fresh information by venturing into unexplored states. Furthermore, the application of VIME to IIEGs characterized by uncertain environments necessitates the development of appropriate interaction strategies for extracting information from the environment. However, the traditional \emph{Counterfactual Regret Minimization} algorithm is typically designed to compute average strategies for players when environmental information is available. Consequently, the design and implementation of effective interaction strategies emerge as a pivotal pathway toward resolving NE in IIEGs with unknown environments.

\textbf{Main contributions.} To solve imperfect information extensive-form games with unknown environments, we propose a method called VCFR. VCFR utilizes \emph{variational information maximizing exploration} for the first time in 2-player zero-sum (2p0s) games to explore uncertain environments. We start by obtaining the posterior distribution of the information by interacting with the real environment, and then we use a Bayesian neural network (i.e., BNN) to model the unknown environment. To increase exploration efficiency, we use information gain to direct the agent towards more significant environmental uncertainty areas. Furthermore, we incorporate information gain into \emph{counterfactual regret minimization} to modify its computation indirectly, resulting in a policy that can be used for finding a NE and as a curiosity-driven interaction strategy. More precisely, VCFR is a plug-and-play algorithm that can be replaced by any CFR-like algorithm, such as \emph{CFR+}\cite{tammelin2015solving}, \emph{DCFR}\cite{brown2019solving}, etc. Finally, we conduct experiments on the standard testbed of 2p0s games with unknown environments. The results show that our method outperforms other baselines by achieving better strategies with fewer interaction rounds.

\section{Related work}\label{sec2}

\subsection{Measure of uncertainty} Various methods have been proposed to address the problem of uncertainty measurement in unknown environments. One such method is the random prior function \cite{osband2018randomized}, originally used to improve the performance of \emph{Bootstrapped DQN}. In this method, bootstrapped functions are trained to fit value $Q$ with a posterior probability, and a fixed prior is given to each function through a random network. Initializing the prior probabilities randomly enhances the diversity of bootstrapped functions, thereby improving the posterior probability distribution. The effectiveness of the uncertainty measurement obtained by fitting a random prior has been proven in both theory and application \cite{DBLP:conf/iclr/BurdaESK19,DBLP:conf/iclr/CiosekFTHT20}. Another widely adopted method for measuring uncertainty is the deep ensembles \cite{lakshminarayanan2016simple}. Each model in the ensemble is trained based on bootstrap data so that the predicted variance between models can be utilized as a measure of epistemic uncertainty. The disadvantage of deep ensembles is that they tend to give overconfident estimates of uncertainty. Moreover, dropout \cite{gal2016dropout, dropout2014} was proposed as a practical tool for modeling uncertainty in deep learning for obtaining uncertainty estimates, and it can be extended to quasi-KL \cite{gal2017concrete}. Hypernet \cite{DBLP:journals/corr/abs-1711-01297} is a network that learns the weights of another network to directly provide parameter uncertainty values, but is shown to be computationally very expensive. In this work, we focus on measuring the uncertainty of the environment with a BNN \cite{brosse2018promises,rezende2014stochastic}. BNN is a traditional approach that combines probabilistic modeling with neural networks and can provide a degree of confidence in prediction results.

\subsection{Exploration methods in reinforcement learning}
The exploration methods in RL for unknown environments can be categorized into the following three main types. The first is optimistic exploration, which is widely used in RL. \emph{Upper-Confidence-Bound} (UCB) exploration \cite{carpentier2011upper}, as used in AlphaGo, is similar to greedy selection, as both favor the latest or best actions. The widely used Deep Q-learning also employs a greedy strategy, known as the $\epsilon$-greedy \cite{DBLP:journals/nature/MnihKSRVBGRFOPB15, https://doi.org/10.1049/trit.2020.0024}, to explore more information in the environment. It randomly selects an action from all possible actions with a probability of $\epsilon$, and with a probability of $1-\epsilon$, it selects $a_{t} = \max_{a}Q(S_{t}, a; \omega)$. The optimistic initial value \cite{shojaee2017optimistic} realizes exploration by increasing the initial value of the function, which essentially explores the state with a lower frequency of occurrence. It is worth noting that the selection of initial values requires prior knowledge, and the exploration will be unstable during the initial stage. Agents tend to select actions with higher entropy values in the gradient bandit algorithm \cite{silver2014deterministic}, where the entropy of each action is adjusted by the rewards. In real-world applications, the idea of optimistic exploration is also adopted by both \emph{asynchronous advantage actor-critic} \cite{https://doi.org/10.1049/rpg2.12782} and \emph{expert actor-based soft actor-critic} (E-SAC) \cite{https://doi.org/10.1049/cit2.12195}. They achieve this by promoting policy exploration through maximum entropy, ensuring that extremely high probabilities are not assigned to any single action within the action space. This strategy helps prevent repetitive selection of the same action, thereby mitigating the risk of getting stuck in local optima.

The second category is posterior sampling \cite{osband2017posterior,chapelle2011empirical}, which incorporates ideas from Bayesian learning and focuses on using posterior probabilities for more targeted exploration. The algorithm based on posterior sampling will modify its probability distribution after each sampling, and the variance of each action is reduced through a large number of samples.

The third category, which is relevant to our work, is exploration based on information gain \cite{russo2014learning}. Information gain is generally comprehended as the intrinsic reward of agents, which can measure the contribution of a new state to information. To reach a state where more rewards can be obtained, an agent selects the actions of maximizing empowerment that is calculated by mutual information \cite{mohamed2015variational}. If a set of states share the same optimal action, then the action can be interpreted as a representation of the states. Another approach \cite{still2012information} also uses mutual information as an exploration reward, aiming to find the action with the most state information among the strategies with uniform rewards by minimizing the mutual information of actions and states.

\section{Preliminaries}
This section briefly introduces the definition of two-player zero-sum imperfect-information extensive games with unknown environments, which is the setting used in our experiment. In addition, we review related methods, namely VIME and CFR, which give us inspiration for a solution.

\subsection{Extensive-form games}
The extensive-form games with two players are special cases of general extensive games with imperfect information, which are usually used to model the sequential decision-making game. Therefore, we first introduce the concept of extensive games (for a complete treatment, see \cite{DBLP:conf/iclr/ZhouLZ20, tammelin2015solving}).

There is a finite set $\mathcal{P}$ of players in an imperfect-information extensive-form game, $\mathcal{P}=\{1,2, \ldots, n\}$. The ``nature'' of the game is a chance player $\mathcal{C}$, which chooses actions with an unknown probability distribution. Here, we define $c^{*}$ as $\mathcal{C}$'s probability of strategies. A history (or state) $h$ represents the sequence of actions taken by a player starting from the initial game state. $H$ is a finite set of possible histories, including the empty sequence $\emptyset$, and $H$ can be thought of as the finite set of all nodes in the game tree. $Z \subseteq H$ refers to the set of all terminal states. For any $z \in Z$, $u_{i}(z)$ is the payoff function for player $i$ when the game ends in state $z$. The set of available actions after a non-terminal history $h$ is referred to $A(h)=\{a: (h,a) \in H\}$. $\mathcal{P}(h)$ is the player who takes an action after the history $h$. In particular, if $\mathcal{P}(h)=\mathcal{C}$, then the chance player chooses an action with a probability after $h$. Let $I_{i}$ be an information partition of all of the states with player $i$ to act. The information set of player $i$ is $\mathcal{I} \in I_{i}$. For any pair of states $h_{1}, h_{2} \in \mathcal{I}$, player $i$ cannot distinguish between them. A strategy profile $\sigma=\left\{\sigma_{1}, \sigma_{2},\ldots, \sigma_{n}\right\}$ is a set of strategies for all players. $\sigma_{i}\left(h_{1}\right)$ and $\sigma_{i}\left(h_{2}\right)$ are equal for player $i$ when $h_{1}, h_{2} \in \mathcal{I} $. Let $r^{*}$ denote the reward function at terminal state, i.e., $r^{*}(h, i)$ is the distribution of the reward of player $i$. The maximal size of available actions for $h$ is referred to $\mathcal{A}=\max _{h}|A(h)|$. 

If action $a \in A(h)$ results in the transition from state $h$ to $h^{\prime}$, then we express this as $h \cdot a = h^{\prime}$.
$\pi^{\sigma}(h)=\Pi_{h^{\prime} \cdot a \sqsubseteq h} \sigma_{P\left(h^{\prime}\right)}\left(h^{\prime}, a\right)$ denotes the probability of reaching $h$ when all players choose actions according to $\sigma$. $ d^{*}=(r^{*},c^{*}) $ can be regarded as the unknown environment, where $r^{*}$ and $c^{*}$ follow a prior distribution $\mathbb{P}_{0}$. This means that the underlying distribution $d^{\ast}$ is sampled from $\mathbb{P}_{0}(c,r)$. Here, $c$ and $r$ are not necessarily independent. Since $c^{*}$ is uncertain under RL, the probability of reaching $h$ depends on $\sigma$ and $c^{*}$. After playing $t$ games, players collect some samples from $d^{\ast}$ and they can obtain the posterior distribution, denoted as $\mathbb{P}_{t}$. Defining by formal formula is $\pi_{\sigma}\left(h \mid d^{*}\right) = \prod_{i \in\{\mathcal{P}\} \cup\{\mathcal{C}\}} \pi_{\sigma}^{i}\left(h \mid (r^{*},c^{*})\right)$. Similarly, we use $v_{i}\left(h \mid \sigma, d^{*}\right)$ to refer to the expected payoff for player $i$ according to $\sigma$.

\subsubsection{Nash equilibrium and exploitability}
Finding an approximate NE \cite{nash1950equilibrium,DBLP:journals/cacm/Roughgarden10} is a significant solution for IIEGs. For convenience, we define $\sigma_{-i}$ as a strategy profile of every player except player $i$. The best response means the strategy $BR(\sigma_{-i})$ that maximizes the reward of player $i$ when $\sigma_{-i}$ is given. That is, $v_{i}(BR(\sigma _{-i}),\sigma _{-i})=\max_{\sigma_{i}^{\prime}}v_{i}(\sigma_{i}^{\prime},\sigma_{-i})$. $\hat{\sigma}=(\hat{\sigma}_{1},\hat{\sigma}_{2})$ is an NE if and only if $v_{i}(\hat{\sigma}\cond d^{\ast}) = \max_{\sigma^{\ast}_{i}}v_{i}(\sigma^{\ast}_{i},\hat{\sigma}_{-i}\cond d^{\ast})$. More specifically, NE has been proven to exist in all finite games. The exploitability $expl(\sigma_{i}\cond d^{\ast})$ assesses how much worse $\sigma_{i}$ performs against $BR(\sigma)$ relative to how an equilibrium strategy $\hat{\sigma}_{i}$ performs against $BR(\hat{\sigma}_{i})$. It is generally used to measure the approximation error of $\hat{\sigma}=(\hat{\sigma}_{1},\hat{\sigma}_{2})$:
\begin{equation}
    % expl(\sigma_{i} \cond d^{*}) = \sum_{\sigma_{i=1,2}} \max_{\sigma^{*}_{i}} v^{i}\left(\sigma^{*}_{-i}, \sigma_{i} \cond d^{*}\right)
    expl(\hat{\sigma} \cond d^{\ast })=\max_{\sigma^{\ast }_{1}} v_{1}(\sigma^{\ast }_{1}, \hat{\sigma }_{2} \cond d^{\ast } )+\max_{\sigma^{\ast }_{2}} v_{2}(\hat{\sigma }_{1}, \sigma^{\ast }_{2} \cond d^{\ast })
\end{equation}

\subsection{Variational information maximizing exploration}
VIME is an exploration algorithm based on the maximization of information gain for uncertain environments in RL problems. It aims to explore the direction in the environment that exhibits the highest uncertainty, thereby acquiring as much information as possible.

VIME adopts a BNN \cite{graves2011practical,blundell2015weight} to model the environment. The information gain used by VIME refers to the reduction degree of information complexity in an environment. It is calculated in the dynamic model of environment, $p\left(s_{t+1} \mid s_{t}, a_{t} ; \theta\right)$, parametrized by the random variable $\theta$. Assuming a prior $p(\theta)$, it sustains a distribution over the dynamic models based on the distribution over $\theta$, and this distribution is updated in a Bayesian manner. Thus, the agent is encouraged to take actions that maximize the reduction in uncertainty. The process of reducing uncertainty can be abstracted as minimizing entropy over a sequence of actions $a_{t}$:
\begin{equation}
    \sum_{t} \left(H\left(\Theta \mid h_{t}, a_{t}\right)-H\left(\Theta \mid S_{t+1}, h_{t}, a_{t}\right)\right),
\end{equation}
where $h_{t}=\{s_{1},a_{1}, \cdots, s_{t}\}$ is the history of agent and $\Theta$ is a set of the random variables $\theta \in \Theta$ about the agent in the environment. In information theory, the individual term is equivalent to the mutual information $\Delta H$ between the distribution of the next state $s_{t+1}$ and the model parameter $\Theta$. Specifically, the mutual information can be expressed by the following formula:
\begin{equation}
\label{eq2}
\Delta H\left(s_{t+1}, \Theta \mid s_{t},a_{t}\right)=H\left(\theta \mid h_{t}\right)-H\left(\theta \mid s_{t},a_{t},s_{t+1}\right),
\end{equation}
which is equal to the Kullback-Leibler (KL) divergence $D_{KL}$. An agent is encouraged to act towards node with greater $D_{KL}$, hence KL divergence can be considered to be consistent with the information gain. If the entropy of $\theta$ can be decreased when the agent is in the state $s_{t+1}$, it indicates that the state $s_{t+1}$ contributes to enhancing the dynamic belief. For an agent, $D_{KL}$ can be interpreted as an intrinsic reward, distinct from external rewards in the environment. Hence, based on the Eq. \ref{eq2}, the reward of the next state can be expressed as:
\begin{equation}
    r^{\prime}(s_{t+1})=r(s_{t})+\eta D_{\mathrm{KL}}\left[p\left(\theta \mid h_{t}, a_{t},  s_{t+1}\right) \| p\left(\theta \mid h_{t}\right)\right],
\end{equation}
where $\eta \in \mathbf{R}^{+}$ is a discount factor and contributes to exploration.

\begin{figure*}[t]
\centering
\includegraphics[width=0.9\textwidth]{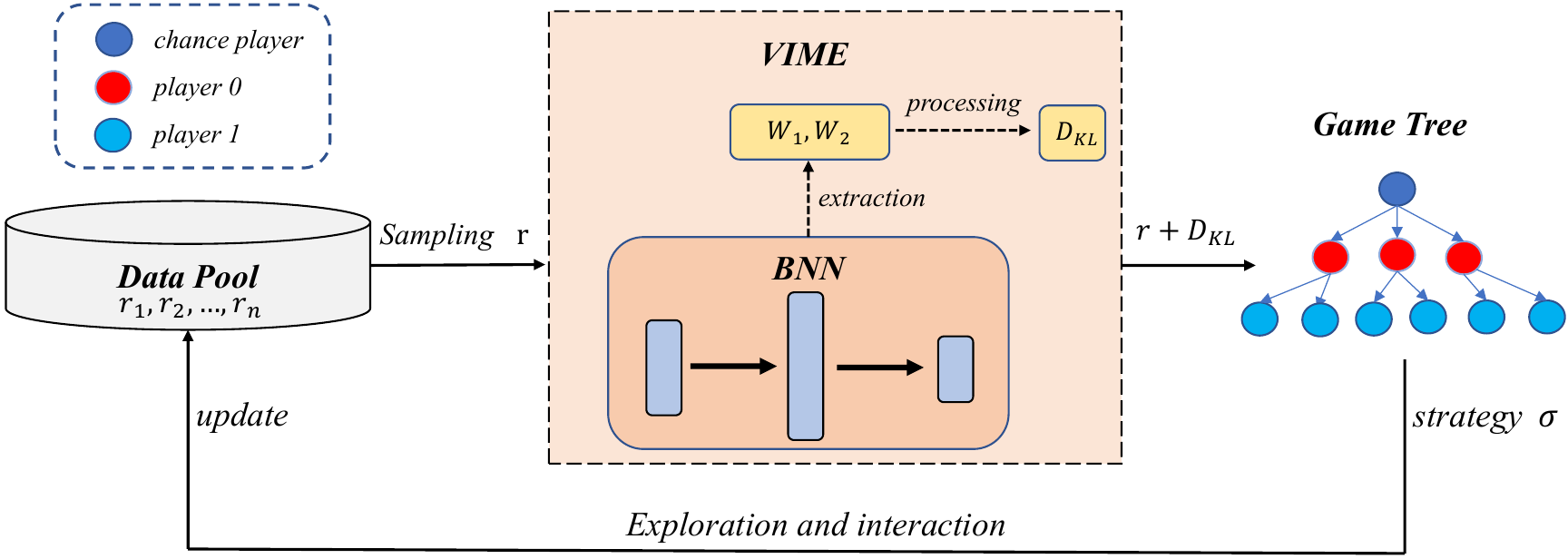}
\caption{An architecture illustration of our proposed method. The model starts by initializing the data pool with the distribution of rewards. The data pool stores the posterior distribution of the reward corresponding to each action. Then the KL divergence for each node can be calculated using the weight distribution of BNN. The reward $r^{\prime}$ that increases the information gain is assigned to the terminal node in the game tree. Finally, the data pool is updated by interacting with the environment using average strategies.}
\label{fig1_1}
\end{figure*}

\subsection{Counterfactual regret minimization}

The CFR algorithm \cite{zinkevich2007regret,brown2019deep}, which converges to NE by constantly iterating to reduce regrets, has been proven to be successful in two-player zero-sum games with incomplete information through experiments. The core idea of CFR is to apply the regret matching algorithms \cite{LITTLESTONE1994212,chaudhuri2009parameter} to each information set for computing strategy profiles. In other words, it divides the total regret into a number of regrets on information sets. The purpose of optimizing policy and finding a NE is achieved by minimizing cumulative regrets. We denote the set of all terminal nodes with a prefix in $\infoset$ as $Z_{\infoset}$, and refer to the particular prefix as $z[\infoset]$. Let $v^{\sigma}_{i}(\infoset)$ be the counterfactual value of player $i\in P(\infoset)$ at the information set $\infoset$:
\begin{equation}
\label{eq5}
v^{\sigma}_{i}(\infoset)=\sum_{z \in Z_{\infoset}} \pi_{-i}^{\sigma}(z[\infoset]) \pi^{\sigma}(z[\infoset] \rightarrow z) u_{i}(z).
\end{equation}
Eq. \ref{eq5} represents the expected payoff to player $i$ when reaching $\infoset$, where $\pi_{-i}^{\sigma}(z[\infoset])$ denotes the probability of players, excluding player $i$, reaching the information set $\infoset$.
The immediate counterfactual regret $r^{t}(\infoset,a)$ is the counterfactual value difference between taking action $a$ and whole the information set $\infoset$ on round $t$:
\begin{equation}
r^{t}(\infoset, a)=v^{\sigma^{t}}(\infoset, a)-v^{\sigma^{t}}(\infoset).
\end{equation}
For an information set $\infoset$, the total counterfactual regret $R$ of action $a$ after iteration $T$ is 
\begin{equation}
R^{T}(\infoset,a)=r^{T}(\infoset,a\cond d^{*})+R^{T-1}(\infoset,a\cond d^{*}),
\end{equation}
where $R^{T-1}(\infoset,a\cond d^{*})$ represents the cumulative regret from the previous round. When $T=1$, $R^{T}(\infoset,a \cond d^{*})=r^{T}(\infoset,a \cond d^{*})$.
Formally, The update of strategy $\sigma^{T+1}$ on round $T+1$ follows as
\begin{equation}
\sigma^{t+1}(\infoset, a\cond d^{*})=\frac{R_{+}^{t}(\infoset, a\mid d^{*})}{\sum_{a^{\prime} \in A(\infoset)} R_{+}^{t}\left(\infoset, a^{\prime}\cond d^{*}\right)},
\end{equation}
where $R^{T}_{+}(\infoset,a \cond d^{*})$ means regret taking only non-negative value. If $\sum_{a^{\prime} \in A(\infoset)}$ $R_{+}^{t}\left(\infoset, a^{\prime} \cond d^{*}\right) \leq 0$, a player will choose a strategy uniformly randomly with probability. And the average $\bar{\sigma}_{p}^{T}(\infoset \cond d^{*})$ for each information set $\infoset$ on iteration $T$ is $\bar{\sigma}_{p}^{T}(\infoset \cond d^{*})=\frac{\sum_{t=1}^{T}\left(\pi_{p}^{\sigma^{t}}(\infoset \cond d^{*}) \sigma_{p}^{t}(\infoset \cond d^{*})\right)}{\sum_{t=1}^{T} \pi_{p}^{\sigma}(\infoset \cond d^{*})}$.

\section{Method}

Our motivation is to propose a more effective exploration method and incorporate information gain into the generation of interactive strategies, aiming to find an approximate NE in fewer interaction rounds. In this section, we provide a detailed description of the proposed VCFR algorithm, which consists of two parts \cite{qiu2021combining}. The first part introduces an exploration method that models unknown environments using a BNN and relies on information gain. The second part describes the adoption of the CFR algorithm with information gain to generate interactive strategies and solve approximate NE. The pseudo-code of our method is presented in detail in Algorithm \ref{algorithm1}.

\subsection{Modeling Environment with BNN}
The process of constructing the environmental model comprises two main steps. Firstly, the posterior distribution of the unknown environment is obtained by interacting with the real environment. Secondly, based on the data acquired in the previous step, we model the unknown environment using a BNN. Unlike conventional neural networks, BNN is capable of modeling the posterior distribution to represent weights in the form of a distribution. This is achieved by introducing uncertainty into the weights of the neural network, providing a regularization effect. BNN propagates the uncertainty of the weights into the prediction process, enabling the generation of confidence in the prediction results. The output of a BNN describes the likelihood of probability distributions, and a posterior distribution can be calculated through sampling or variational inference. BNNs possess the ability to quantify uncertain information and exhibit strong robustness, making them highly suitable for modeling environmental tasks.

\begin{algorithm}
\caption{VCFR}\label{algorithm1}
\begin{algorithmic}
\While{$t < T$}
\State Sample $d_{t}$ according to observed rewards and transition probability from an unknown environment
\For {each update}
\State Sample the posterior distribution of rewards from $\mathbb{P}_{t}(r\cond z)$, $z\in Z$
\State Collect all $(z,r)$ as datasets
    \For{each $(z,r)$ in datasets}
    \State Calculate information gain by Eq.\ref{eq9} $ \begin{cases} \mbox{maintain the update frequency}, & D_{KL}>\lambda \\ \mbox{reduce the update frequency}, & D_{KL}<\lambda \end{cases}$
    \State Construct a new reward based on the information gain $r^{\prime}=r+\eta D_{KL}$
    \EndFor
\State Train BNN and update the distribution of the environment
\EndFor
\For { $i\in \{1,2\}$ }
\For {$h\in H$}
\If {the probability that node arrives $probs<1e-5$}
   \State prune this node
\EndIf
\State Calculate counterfactual regrets with \emph{Regret Matching}
\EndFor
\State Calculate the average strategy $\bar{\sigma}_{i}^{T}$ with information gain.
\EndFor
\ForAll{$i\in \{1,2\}$}
\State Gather the environment data and update Observed rewards and transition probability with interaction strategy $\bar{\sigma}_{i}^{T}$
\EndFor
\EndWhile
\end{algorithmic}
\end{algorithm}

\begin{figure}[ht]
\centering
\includegraphics[width=0.8\linewidth]{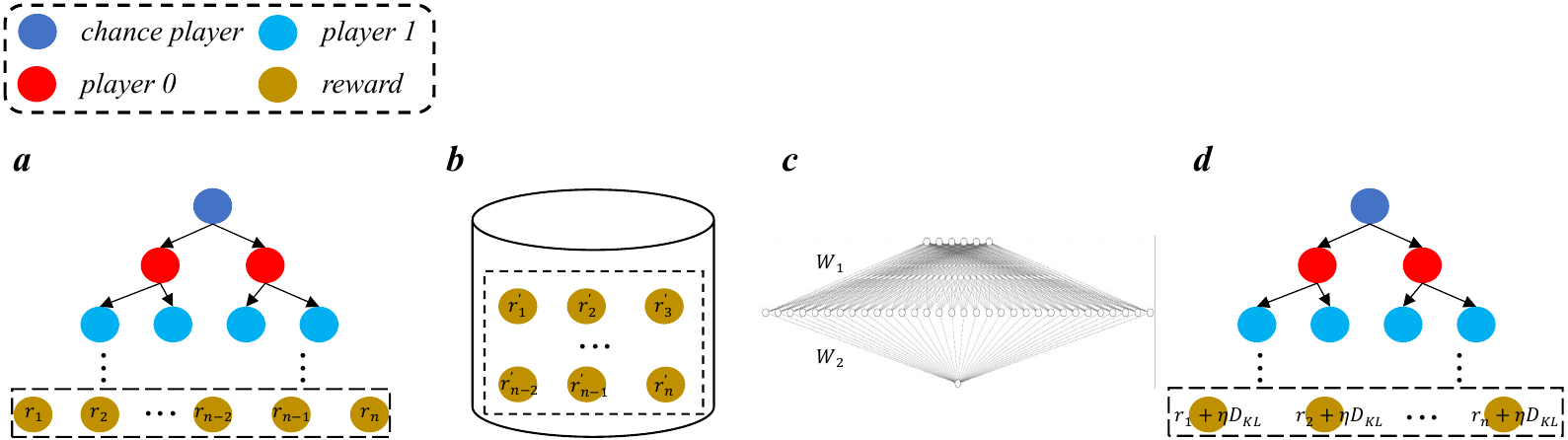} % Reduce the figure size so that it is slightly narrower than the column. Don't use precise values for figure width.This setup will avoid overfull boxes.
\caption{A flowchart of our method. \textbf{a}: Constructing a complete game tree and performing \emph{dirichlet} sampling on rewards in the real game environment. The accuracy of sampling improves progressively with continuous updates to the environment. $r_{n}$ represents the sampled reward.
\textbf{b}: The data in the data pool primarily consists of posterior distributions of rewards corresponding to each action. $r_{n}^{\prime}$ represents the posterior distribution of reward.
\textbf{c}: It shows the BNN used for our work. The one-hot encoding of nodes is input into the BNN, producing posterior samples of rewards. $W_{1}$ and $W_{2}$ represent the weight distributions in the BNN. The KL divergence value for each node in the game tree is obtained through additional computation on the weights of the BNN.
\textbf{d}: The value of KL divergence is considered equivalent to information gain, and it is added as the agent's intrinsic reward to the original reward (i.e., $r_{n}+\eta D_{KL}$). This facilitates the agent in generating strategies to continue exploring unknown environments and find an NE.}
\label{fig1_2}
\end{figure}

The architecture of our proposed method is shown in Figure \ref{fig1_1}, and the flowchart is depicted in Figure \ref{fig1_2}. The data pool $\mathbb{L}$ stores the posterior distribution of rewards corresponding to each action. CFR can calculate the average strategy with the new reward $r^{\prime}$ added with information gain and explore the environment to collect the data according to the curiosity-driven strategy. The approximate NE will be found after continuous iterations. In this work, BNN maintains the players' dynamics model, denoted $p(r\cond d_{t},z;\theta)$, where $d_{t}$ represents the sampled environment. Precisely, the BNN consists of three layers. The parameters of the input and hidden layers are initialized as \emph{Gaussian} distributions, and the \emph{ReLU} is applied to the first two layers as the activation function. The identity function serves as the activation function for the output layer. In terms of network architecture, the input layer has a size of 6, the hidden layer is of size 32, and the output layer has a size of 1. In this work, the BNN takes the one-hot encoding of nodes as input, representing a 6-dimensional vector for terminal nodes, as rewards are distributed only at the terminal nodes. More precisely, we perform one-hot encoding and then use \emph{Principal Component Analysis} (PCA) to reduce the dimensionality to 6. The produced output of the BNN is posterior samples of rewards. The KL divergence value for each node in the game tree is obtained through additional computation on the weights of the BNN. The batch size is either 500 or 1000, depending on the scale of the game scenario. The learning rate used for training is set to 0.005. Furthermore, we employ the \emph{Adam} optimizer to update the parameters.

However, even with a modeled environment, finding an approximate NE remains difficult without an effective method for updating and exploring it. To make the modeled unknown environment closer to the real environment and enhance the efficiency of exploration strategies, we employ information gain for more targeted exploration. The information gain measures the contribution of each sample by the BNN to the environmental update in a single sampling. It quantifies the difference between the distributions before and after the update, which can be described using the KL divergence between the two distributions:
\begin{equation}\label{eq9}
I\left(r ; \Theta \mid d_{t},z\right)=\mathbb{E}_{r}\left[D_{\mathrm{KL}}\left[p\left(\theta \mid d_{t},z\right) \| p\left(\theta \mid d^{*},z\right)\right]\right]
\end{equation}

We take the calculated KL divergence $D_{KL}$ as a measure of the player's desire to explore. In other words, the information gain can be considered numerically equal to the KL divergence. The uncertainty of the environment is treated as an intrinsic reward for a player. We set a threshold $\lambda$ for KL divergence. In order to explore the direction of greater curiosity about the environment, we set the threshold $\lambda$ to 1, and use $\lambda$ to periodically update the value of information gain. The update frequency refers to the number of samples taken at intervals. Normally, information gain is updated with each sampling. The update frequency remains unchanged when $D_{KL}>\lambda$. On the contrary, It indicates that the desire for exploration is low when $D_{KL}<\lambda$, and the update frequency is reduced. The computational efficiency of information gain can be greatly improved without affecting the exploration results. The original reward $r_{t}$ adds to KL divergence for obtaining a new reward $r^{\prime}_{t}$ with information gain as follows:

\begin{equation}
r^{\prime}_{t}=r_{t}+\eta D_{\mathrm{KL}}
\label{equation:10}
\end{equation} 
where $r^{\prime}_{t}$ will be used later as a processed reward on the interaction strategy. The hyperparameter $\eta$ is a key factor in determining the degree of exploration. In our work, we set $\eta$ to 0.005 to strike a balance between the influence on CFR computation results and the effectiveness of exploration. During the learning process, maximizing $r^{\prime}$ enables a balance between exploration and exploitation, facilitating the gradual convergence of the environment to the true one.

\subsection{Information gain based CFR}

The purpose of information gain based CFR is to find an approximate NE and obtain the average strategy. This average strategy can not only be utilized by the player but also be directly used as an interaction strategy to explore the environment. The traditional CFR algorithm continuously minimizes the regret $R^{T}(I,a)$  by inputting the information of the game tree, such as the strategy combination of each node and the reward of the terminal node. However, the real environment $d^{*}$ and the sampled one are different, so it is not possible to effectively reduce the regret value. In other words, the approximate NE of sampled environments will not be eventually found by continuously decreasing the exploitability in a real environment. Formally, the relationship between exploitability and regret can be expressed by the following formula:
\begin{equation}
\begin{aligned}
expl\left(\bar{\sigma}_{i} \mid d^{*}\right)= &\frac{1}{T}\Big(\sum_{i\in \{1,2 \}}R_{i}^{T}+\sum_{t\leq T}\big(u_{i}(\sigma_{T}^{*} \mid d^{*})  -u_{i}(\sigma_{T}^{i} \mid d_{t}) \big)\Big)
\end{aligned}
\end{equation}

We have also made some improvements to the CFR for situations where some environmental information cannot be known. We use the $r^{\prime}$, which adds to information gain, obtained in Equation (\ref{equation:10}) as one of the environmental information sources for CFR. The information gain is added to the terminal node of the game tree. Due to the recursive and iterative characteristics of CFR, the information gain can affect each node from the bottom to the top. Different from traditional CFR, which continuously reduces the exploitability to improve the strategy, the addition of information gain is able to keep the reward $r^{\prime}$ to the direction of environmental exploration to improve the effect. The large cost of time and space is always a difficulty in the problem of extensive games with imperfect information. Inspired by pruning \cite{brown2015regret}, judging its arrival probability first for each node in the game tree. When the node is at an extremely low arrival probability, it will be regarded as a relatively invalid node and not be traversed in this round. For all remaining nodes, the player $i$ makes use of the current strategy $\sigma^{t}_{i}$ to calculate the cumulative regret $R_{t}$ and the counterfactual value (CFV) $v^{\sigma^{t}}_{i}(h)$ \cite{DBLP:conf/iclr/LiHZQS20}. Through the regret matching (RM) \cite{hart2000simple} of regret value for each node, the $\sigma^{t}_{i}$ can be calculated. In the end, a game tree with $v^{\sigma^{t}}_{i}(h)$, $R_{t}$ and $\sigma^{t}_{i}$ for all valid nodes can be obtained. The significance of the average cumulative regret value $\bar{R_{T}}$ and the average strategy $\bar{\sigma}^{T}$  is that there is a non-negligible relationship between them and NE. Despite the unknown environmental information used in our method, our goal is still to minimize $R_{T}$ and improve the player's reward so that the average strategy approaches an approximate NE.

The interaction strategy is a critical factor in determining the environmental certainty in our experimental setup. The convergence of the sampling environment to the real environment significantly affects the exploitability calculation. Since environmental uncertainty is extremely high in the initial state of a randomly initialized environment, there is a greater variance in environmental distribution.  To reduce this variance and converge the unknown environment, we use the average strategies with information gain to interact with the real environment directly for collecting data in our approach.

\section{Experiments}
This section focuses on the details of the experiment, including a description of the representative baselines.  Finally, we present the experimental results and provide an analysis.

\subsection{Experimental setup}
Poker, as a game with imperfect information, provides an ideal testbed for evaluating equilibrium-finding algorithms. In recent years, poker games have been used to verify techniques related to handling imperfect information. In this work, we used variants of Leduc Hold'em Poker \cite{DBLP:journals/corr/abs-1207-1411} and Kuhn Poker \cite{Kuhn+1951+97+104}. More specifically, we introduced modifications to Leduc Hold'em Poker, a game involving two players with predetermined betting and raising amounts. It should be emphasized that while the modified Leduc Hold'em Poker maintains the original game tree structure, the transition probabilities for the chance player $c$ and the reward function $r$ for terminal nodes have become uncertain. Similar adjustments were applied to Kuhn Poker. These adaptations were meticulously executed to establish an experimental environment conducive to the comprehensive assessment of our proposed methodology.

To control the sizes of the generated game tree, each player is constrained to bid no more than four or five times the big blind in the Leduc poker. The game trees generated from Leduc(4) and Leduc(5) contain $9,652$ and $34,438$ nodes, respectively. The reward $r$ is initialized randomly from $\{-1,1\}$, and the reward function $r(h)$ follows a binary distribution.

We take advantage of the BNN to model the environment. The BNN architecture we have adopted is shown in Figure \ref{fig1_2}. The BNN has a depth of three layers. The parameters of the input and hidden layers are set to the Gaussian distribution. The 6-dimensional vector matrix of a single terminal node is encoded as the input to the BNN. We perform 20,000 iterations using a batch size of 500 in Leduc(4) and 1000 in Leduc(5).

\subsection{Baselines}
We choose four kinds of methods as our baselines. The description of baselines is given below:
\begin{itemize}
\item \textbf{PSRLCFR}: The PSRLCFR \cite{DBLP:conf/iclr/ZhouLZ20} method combines the CFR and the posterior sampling for RL. The unknown environment is transformed into a known one through Thompson sampling, and the CFR algorithm is used to calculate the approximate NE. In order to update data in the environment, the special strategies $\hat{\sigma}_{1}$ and $\hat{\sigma}_{2}$  are regarded as interaction strategies.

\begin{figure*}
\centerline{\includegraphics[width=0.75\linewidth]{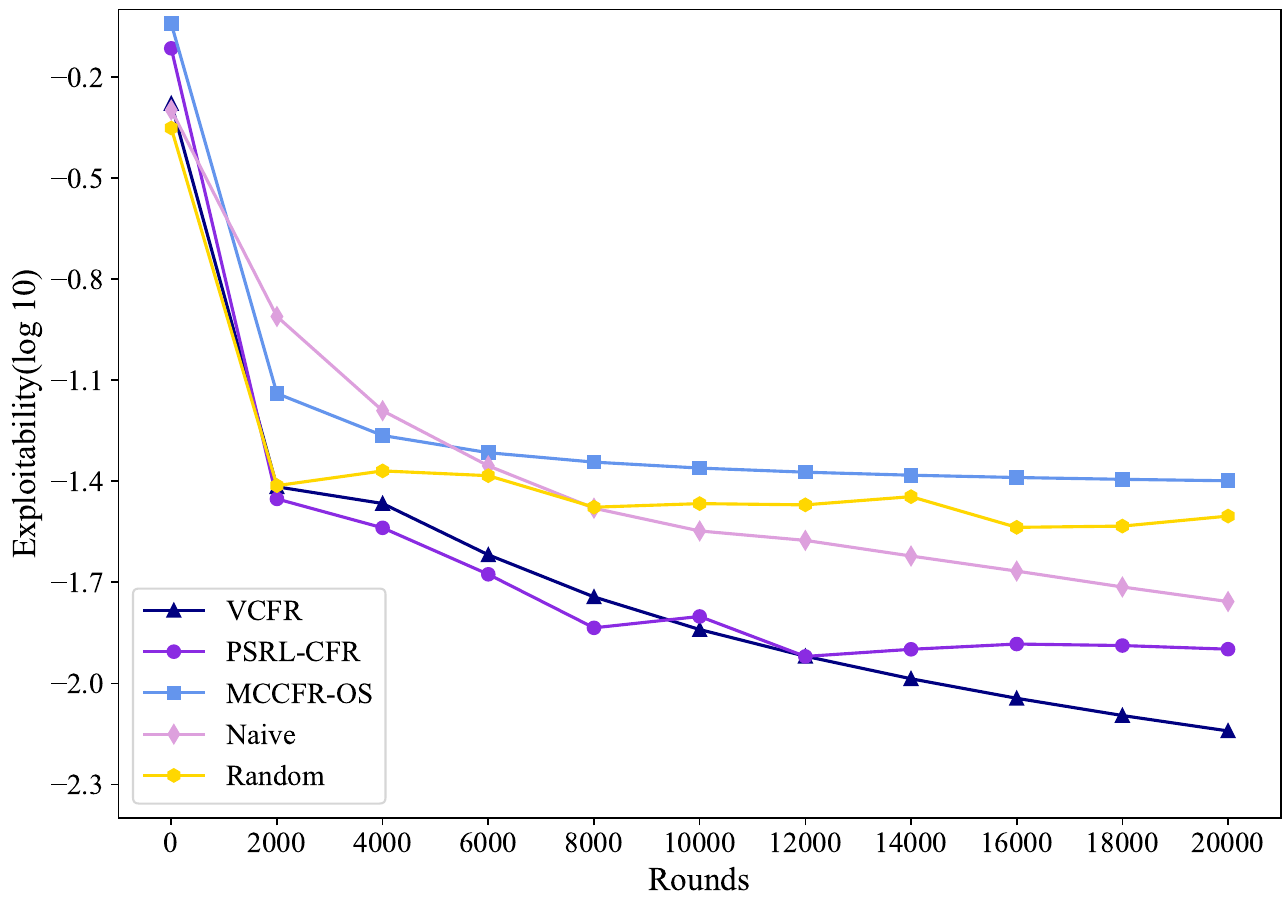}}
\caption{Performance comparison of VCFR and other algorithms in Kuhn poker with unknown environments.} 
\label{Kuhn_vcfr}
\end{figure*}

\begin{figure*}[ht]
\centering
\subfigure[VCFR and other algorithms in Leduc(4)]{
\label{fig2_1}
\includegraphics[width=0.48\textwidth]{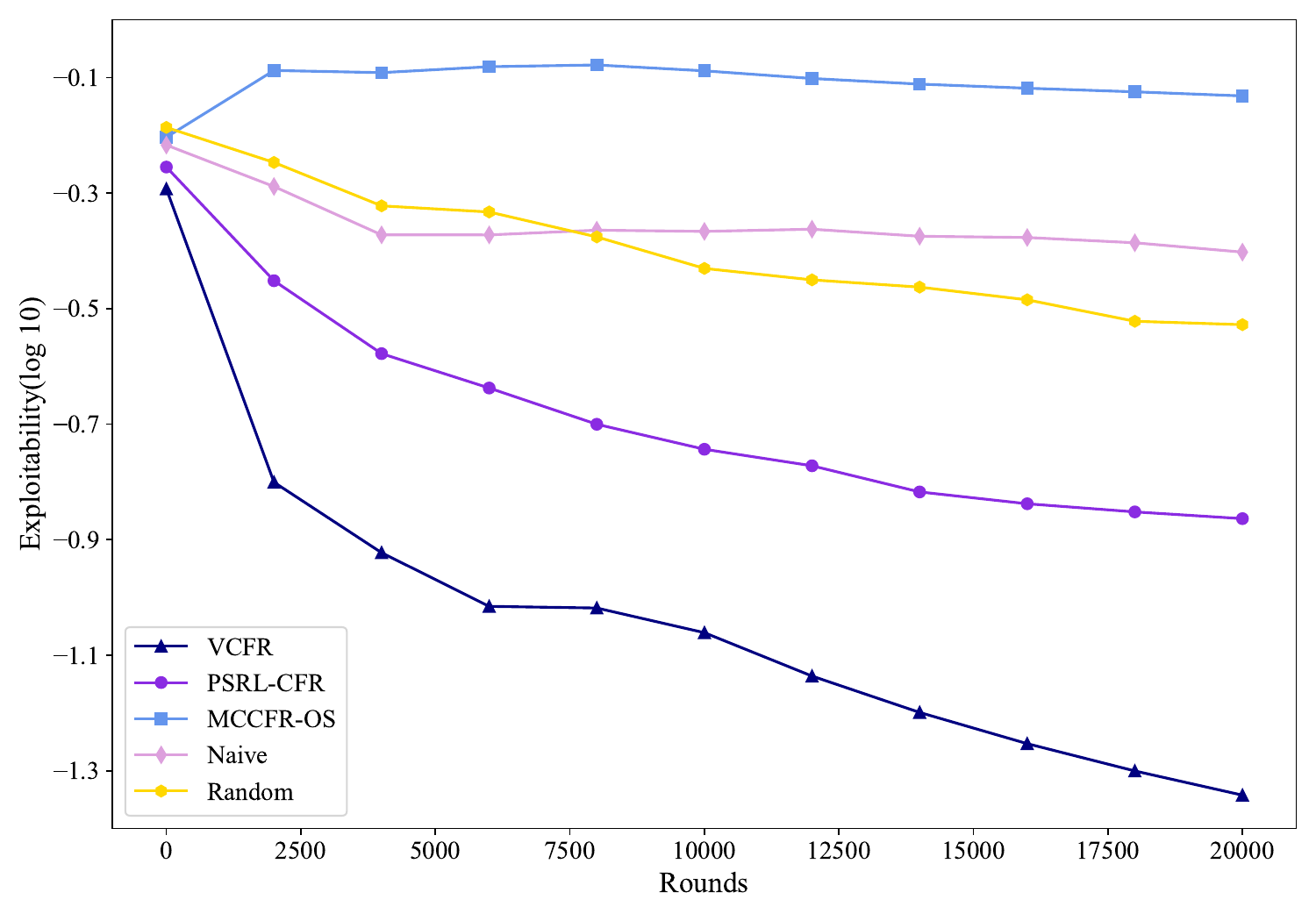}}
\subfigure[VCFR and other algorithms in Leduc(5)]{
\label{fig2_2}
\includegraphics[width=0.48\textwidth]{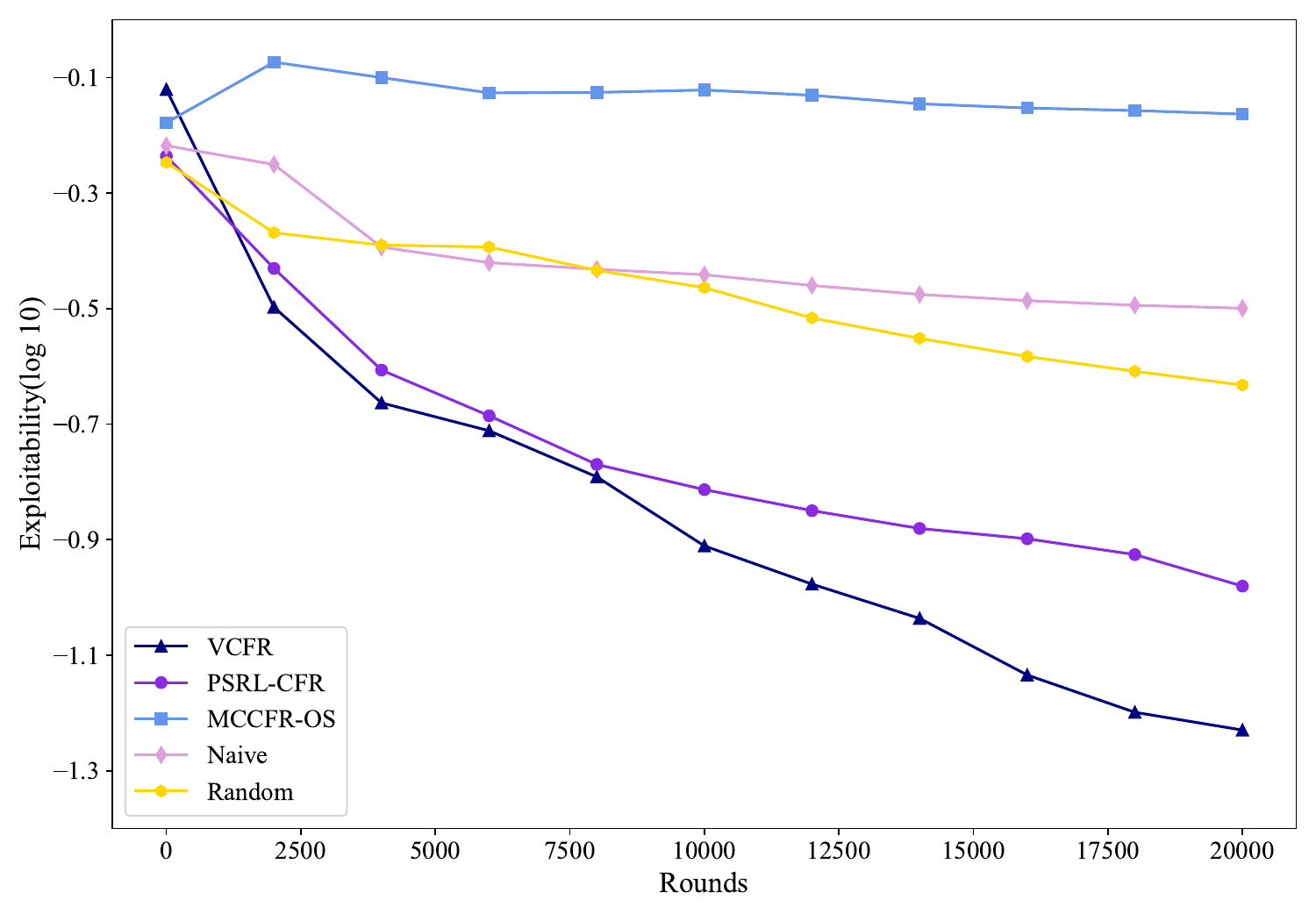}}
 % Reduce the figure size so that it is slightly narrower than the column.
\caption{Comparison between VCFR and other algorithms in different sizes of Leduc poker with unknown environments. Panels~\ref{fig2_1} and~\ref{fig2_2}  respectively depict the experimental results in Leduc(4) and Leduc(5).}
\label{fig2}
\end{figure*}

\begin{figure*}[ht]
\centering
\subfigure[VDCFR in Leduc(4)]{
\label{fig3_1}
\includegraphics[width=0.48\textwidth]{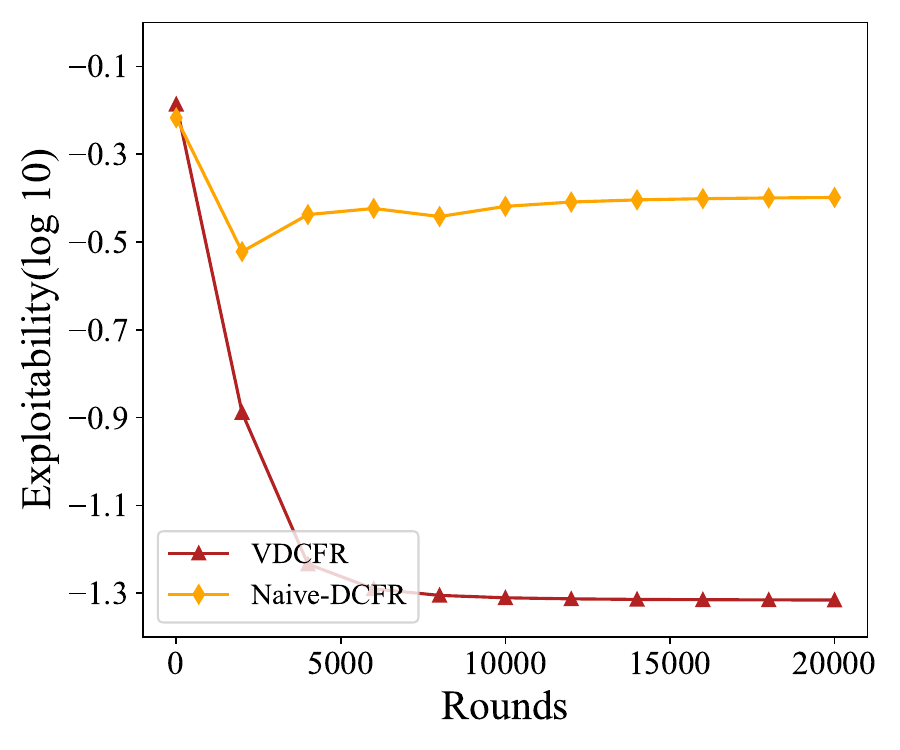}}
\subfigure[VDCFR in Leduc(5)]{
\label{fig3_2}
\includegraphics[width=0.48\textwidth]{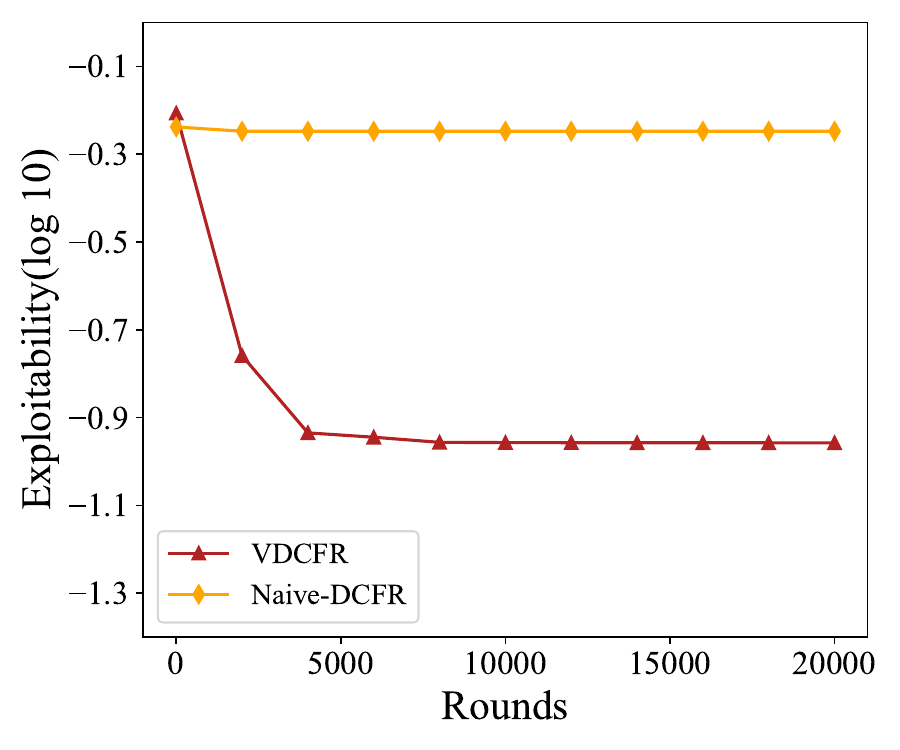}}
\subfigure[VFSP in Leduc(4)]{
\label{fig3_3}
\includegraphics[width=0.48\textwidth]{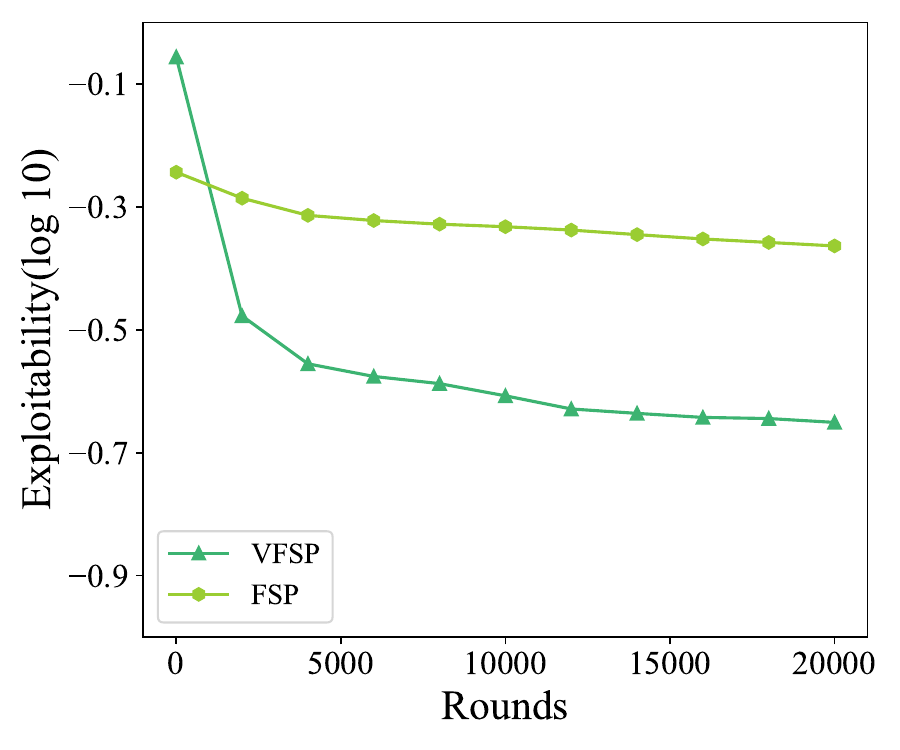}}
\subfigure[VFSP in Leduc(5)]{
\label{fig3_4}
\includegraphics[width=0.48\textwidth]{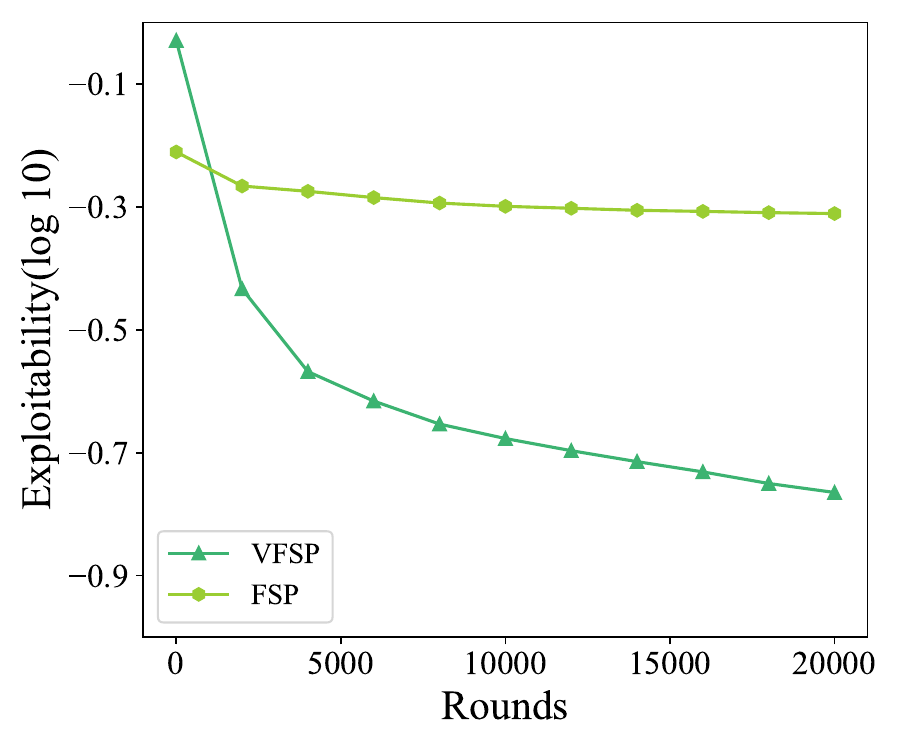}}
\caption{Ablation experiments for our method were conducted on two different scales of Leduc poker with unknown environments. Panels~\ref{fig3_1} and~\ref{fig3_2} present the results of the ablation experiments for VDCFR in Leduc(4) and Leduc(5), respectively. Panels~\ref{fig3_3} and~\ref{fig3_4} illustrate the ablation experiment results for VFSP in Leduc(4) and Leduc(5), respectively.}
\label{fig3}
\end{figure*}

\item \textbf{FSP}: Fictitious self-play (FSP) \cite{DBLP:journals/ml/ConitzerS07,heinrich2015fictitious} is a popular algorithm to find NE in uncertain settings. FSP needs to make use of Fitted-Q iteration \cite{DBLP:conf/iclr/ZhouLZ20} with initial hyperparameters to learn the best response to the average strategy of each player's opponent when the environment is unknown.
\item \textbf{MCCFR-OS}: Monte Carlo counterfactual regret minimization based on outcome sampling \cite{lanctot2009monte} is a popular variant of CFR that avoids traversing the entire game-tree by sampling only a single playing of the game on each iteration. The $\epsilon$-greedy is the exploration strategy where $\epsilon=0.1$.

\item \textbf{Variants of VCFR}: In order to better measure the validity of our experimental methods, we use three additional variants of VCFR: 1) \textbf{Naive}: The reward without information gain is inputted into the process of CFR's computation, and the average strategy calculated is used as exploration policy to interact with the unknown environment; 2) \textbf{Naive-DCFR}: To prove the generalization of our algorithm framework, we use other variants of the Discounted CFR (DCFR) \cite{brown2019solving}. DCFR is a variant algorithm of CFR, and it has three parameters $\alpha$, $\beta$, and $\gamma$ as discount factors to enhance the solution speed. In every round $t$, the effects of three parameters are multiplying cumulative regrets $R$ by $t^{\alpha}/ (t^{\alpha}+1)$ when $R>0$, by $t^{\beta}/(t^{\beta}+1)$ when $R<0$, and contributions to average policy $\bar{\sigma}$ by $\left(t/(t+1)\right)^{\gamma}$. The basic process is the same as Naive-CFR, only replacing CFR with DCFR. The previous experiments had excellent performance when setting $\alpha =\frac{3}{2}$, $\beta = 0$, and $\gamma=2$; 3) \textbf{Random}: The players take random actions in each round.
\end{itemize}

\subsection{Experimental results}

We ran all algorithms for 20,000 iteration rounds on Kuhn poker, Leduc(4), and Leduc(5), with each iteration involving an interaction with the environment to compute the strategy. Exploitability, a commonly used metric in IIEGs, was employed as the evaluation criterion. In the case of the least complex Kuhn poker, our method demonstrated a clear advantage over the state-of-the-art algorithm (i.e., PSRLCFR), as shown in Figure \ref{Kuhn_vcfr}. The fluctuating curves of PSRLCFR algorithm in Figure \ref{Kuhn_vcfr} can be attributed to the low efficiency of its single-sample sampling, leading to unstable convergence of the environment. As exploitability continuously decreased, it indicated the effectiveness of our method, and the strategies obtained within the same number of rounds were closer to NE.

Figure \ref{fig2} presents a performance comparison of VCFR and several baselines in Leduc poker with varying scales. We can see that our proposed VCFR method performs better than other algorithms in two different experimental settings. Specifically, in the smaller-scale Leduc(4), VCFR achieved an exploitability of -0.135 after 20,000 rounds. The algorithm with the second-best performance was PSRLCFR, with an exploitability of -0.864. Although both Random and PSRLCFR demonstrated decreasing exploitability, their convergence speed and lower bounds were inferior to our method. The comparison between Naive CFR and VCFR highlights that the incorporation of information gain in the average strategy not only accelerates strategy convergence but also effectively explores unknown environments.  

In Leduc(5) with higher space complexity, the results in Figure \ref{fig2_2} demonstrate that our method still maintains excellent performance. After 20,000 iterations, the exploitability of VCFR, PSRLCFR, Random, Naive, and MCCFR-OS are -1.230, -0.980, -0.632, -0.499 and -0.163, respectively. However, in Leduc(5), the rate of exploitability reduction for VCFR is slower compared to that in Leduc(4). This might be attributed to the decrease in the convergence speed of the BNN modeling environment with the increasing complexity of the environment. MCCFR-OS and Random exhibit poor performance in both game environments, likely stemming from the inefficiency of their exploration strategies.

\begin{figure*}[ht]
    \centering
    \subfigure[VDCFR in Kuhn]{
    \label{fig4_1}
    \includegraphics[width=0.48\linewidth]{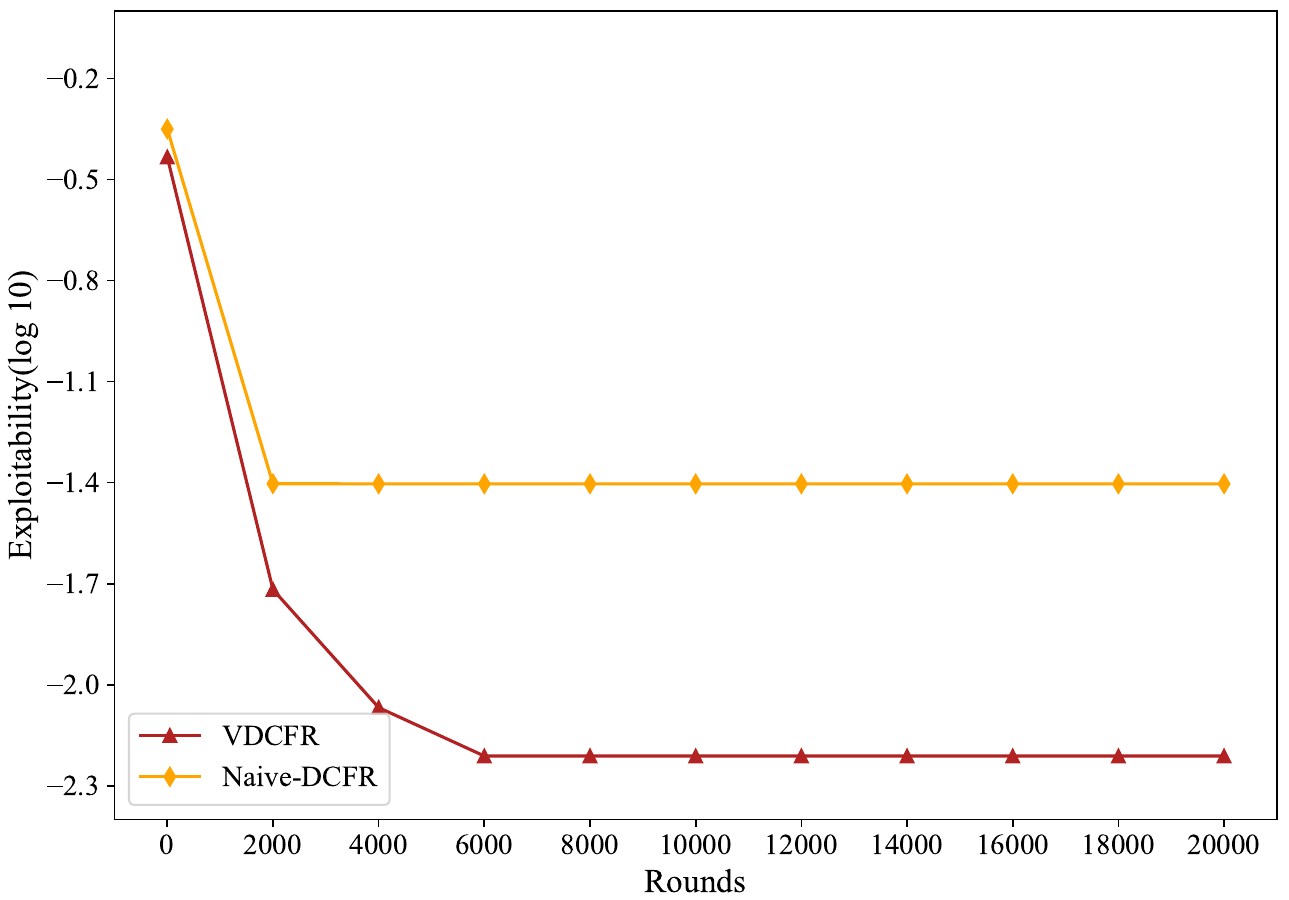}}
    \subfigure[VFSP in Kuhn]{
    \label{fig4_2}
    \includegraphics[width=0.48\linewidth]{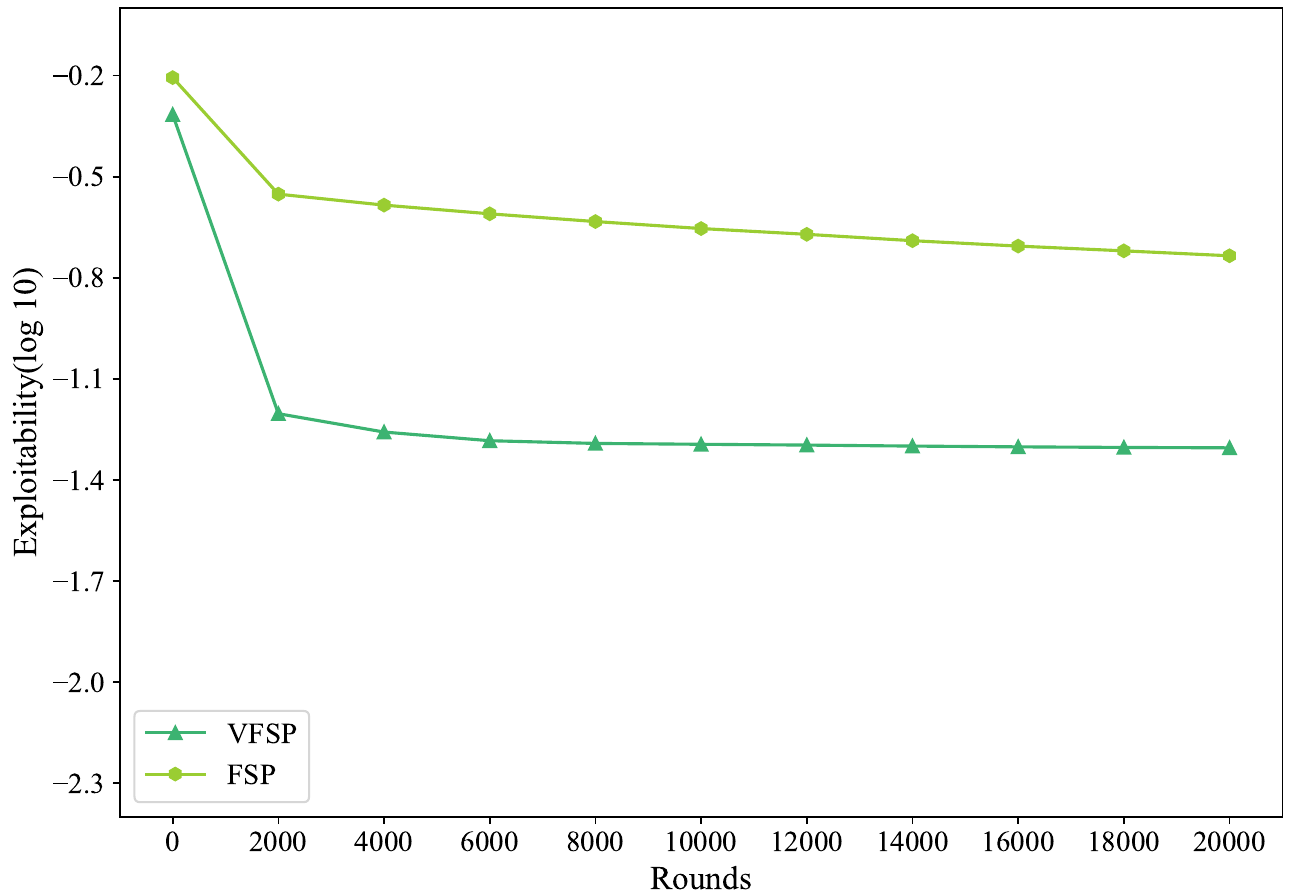}}
    \caption{The ablation experiments of our methods, VDCFR and VFSP, were conducted in Kuhn poker with unknown environments.}
\label{kuhn_VDCFR_VFSP}
\end{figure*}

To further validate the superior generalization ability of the proposed method, we introduced two additional algorithms: VDCFR and VFSP. The architectures of VDCFR and VFSP closely resemble that of VCFR, with the only difference being the replacement of CFR with DCFR or FSP.

We conducted six ablation experiments in three different scenarios, as shown in Figure \ref{fig3} and \ref{kuhn_VDCFR_VFSP}. The excellent performance across scenarios of different scales demonstrates the superior generalization capability of our method. By comparing the VFSP and FSP ablation experiments, as well as the VDCFR and Naive-DCFR experiments, we can observe the significant influence of information gain on the average strategy, enabling effective exploration of the environment. Figure \ref{fig3_1}-\ref{fig3_4} show that the algorithm with information gain can speed up finding approximate NE, and the interaction strategy promotes convergence in an unknown environment.

\section{Conclusion and discussion}

We propose an algorithm called VCFR that combines \emph{counterfactual regret minimization} with information gain to achieve efficient exploration strategies for approximating NE in 2p0s games with unknown environments. The modular design of our approach offers greater flexibility, allowing for the replacement of the CFR algorithm with any approximate NE solver. Empirical results demonstrate the superiority of our approach over other baselines.

In future research, there are several opportunities to optimize our approach. Currently, our method entails computing the KL divergence at each round. Although we attempted to set a threshold and reduce the experiment duration, the impact has remained limited. Therefore, the development of a practical approach to reducing the computation time of KL divergence stands as a potential area for improvement. Furthermore, methods to decrease our approach's dependence on the specific structure of two-player zero-sum extensive games are worth exploring. Subsequent work should also extend our approach to other game types, such as extensive form games involving three or more players.

%\backmatter

%%%%%%%%%%%%%%%%%%%%%%%%%%%%%%%%%
% \bmsection*{Author contributions}

% This is an author contribution text. This is an author contribution text. This is an author contribution text. This is an author contribution text. This is an author contribution text.

\bmsection*{Data Availability Statement}
The data that support the findings of this study are available from the corresponding author upon reasonable request.

\bmsection*{Financial disclosure}

None reported.

\bmsection*{Conflict of interest}

The authors declare no potential conflict of interests.

\bmsection*{Acknowledgments}
This research was funded in part by the National Natural Science Foundation of China under Grant 62376073, Guangdong Provincial Key Laboratory of Novel Security Intelligence Technologies under Grant 2022B1212010005, and the Colleges and Universities Stable Support Project of Shenzhen under Grant GXWD20220811173149002.

\bibliography{wileyNJDv5_AMA}

\end{document}